\documentclass[11pt,notoc,letterpaper]{JHEP3}

\def\beq{\begin{equation}}
\def\eeq{\end{equation}}
\def\ba{\begin{array}}
\def\ea{\end{array}}
\def\bea{\begin{eqnarray}}
\def\eea{\end{eqnarray}}
\def\gev{\ \mbox{GeV}}
\def\tev{\ \mbox{TeV}}

\def\gam{\gamma\gamma}

\usepackage{axodraw}
\usepackage{epsfig}
\usepackage{graphicx}

\title{Dirac Magnetic Monopole Production from Photon Fusion in Proton Collisions}
\author{Triston Dougall \\
   Department of Physics, \\
   Southern Methodist University \\
   Dallas, TX 75275 }
\author{Stuart D. Wick \\
   Department of Physics, \\
   Southern Methodist University \\
   Dallas, TX 75275 \\
\email{s-wick@northwestern.edu}}

\abstract{ 
We calculate the lowest order cross--section
for Dirac magnetic monopole production from photon fusion ($\gam$)
in $p \bar{p}$ collisions at $\sqrt{s}=1.96$~TeV, 
$p p$ collisions at $\sqrt{s}=14$~TeV, and we compare 
$\gam$ with Drell--Yan (DY) production. 
We find the total $\gam$ cross--section is comparable with DY at $\sqrt{s}=1.96$~TeV  
and dominates DY by a factor $> 50$ at $\sqrt{s}=14$~TeV. 
We conclude that both the $\gam$ and DY processes allow for a  
monopole mass limit $m>370\gev$ based upon the null results of the 
recent monopole search at the Collider Detector at Fermilab (CDF).  
We also conclude that $\gam$ production is the
leading mechanism to be considered for direct monopole 
searches at the Large Hadron Collider (LHC).
}

\preprint{SMU-HEP-07-12}

\begin{document}

\section{Introduction}
\label{sec:intro}

Magnetic monopoles have been a theoretical curiosity since the founding of
electromagnetic theory and have motivated numerous innovative experimental searches.
Maxwell's equations possess a dual, electric--magnetic symmetry that goes
unrealized without the discovery of magnetic charges.  Stronger motivation
for monopole searches was provided by
Dirac who showed that the existence of a single magnetic monopole is sufficient to explain the
observed quantization of electric charge \cite{Dirac}, an empirical fact
which goes otherwise unexplained.  Magnetic monopoles are also present 
in a majority of grand unified theories (GUT) of particle 
interactions where monopoles are produced as topological defects during the GUT phase 
transition \cite{tHooftPolyakov}.  Consequently, if a GUT were realized in the early 
universe, after an era of inflation, then a population of magnetic monopoles 
would be left over as relics of the Big Bang \cite{Kibble}.

Decades of interest in monopoles has inspired monopole
searches in a wide range of physical settings \cite{Giacomelli}. 
It has been proposed that relativistic monopoles could be observed 
as a component of the cosmic rays \cite{Wick} and recent flux limits 
have been reported \cite{Besson} employing these techniques.  
Monopoles searches have been conducted in exotic materials like moon rocks \cite{moonrocks} 
and terrestrial materials exposed to excessive radiation \cite{Kalb}.   
Collider searches for directly produced monopoles have been performed, most recently
\cite{Mulhearn,CDF06}.  Despite all efforts to date 
there are no definitive signals for the existence of magnetic monopoles.

The quantization of angular momentum for magnetic and electric poles 
yields the Dirac quantization condition, $eg=n/2$ (setting 
$\hbar=c=1$) for electric and magnetic charges $e$ and $g$, respectively.  
The magnetic charge
\beq
g=\frac{1}{2e}=\left(\frac{e}{2\alpha}\right)\simeq (68.5)e
\label{eq:monocharge}
\eeq
is large in units of the electric charge (where we choose $n=1$ and 
define $\alpha \equiv e^2 \simeq 1/137$).  The large monopole charge implies a 
strong monopole--photon coupling, a characteristic of magnetic monopoles 
that will be exploited in this article.

A useful theory of monopole interactions does not currently exist to perform 
direct production calculations.  
The large monopole--photon coupling precludes the use of perturbation theory
leaving us with a lowest order approximation as our only means to proceed.
Previous authors \cite{Kalb,Mulhearn,CDF06,Bauer} have employed 
a minimal model of monopole interactions which 
assumes a monopole--photon coupling that is proportional
to the monopole's induced electric field $g\beta$ for a monopole moving with velocity
$\beta=v/c.$  We follow these authors and use this same minimal model in our photon
fusion ($\gam$) and Drell--Yan (DY) calculations that follow.

Monopole searches at colliders are restricted to Dirac--type monopoles
(and antimonopoles), which are hypothesized to be
fundamental particles dual to the electron (and positron).
GUT monopoles are excluded in collider searches as they generally have 
too large a mass ($M\sim 100 \Lambda_{\rm{GUT}}$ where $\Lambda_{\rm{GUT}}$ is the 
GUT symmetry breaking scale) and their internal structure exponentially 
suppresses their production cross--section. 
Models of Dirac monopole production have relied extensively on the DY process 
in which a quark and antiquark ($q\bar{q}$) from interacting protons annihilate
to produce a monopole--antimonopole pair ($m\bar{m}$). (For an extensive review 
of the Drell--Yan process see \cite{Stroynowski}.)
The CDF Collaboration at the Fermilab Tevatron recently reported the results of a search
for the direct production of magnetic monopoles \cite{CDF06}.
With no monopole events found, CDF sets a mass limit  $m>360$~GeV
assuming DY production of monopoles. 
A particle collider that probes a new energy frontier, 
such as the Large Hadron Collider (LHC), will open up new physics 
possibilities including the potential for the discovery of magnetic monopoles.  
Future monopole searches are likely to be undertaken at the LHC and will require 
detailed simulations of monopole events based upon our best knowledge of 
the leading monopole production mechanisms.  
Anticipating this need we investigate DY production at the LHC and consider 
an alternative production mechanism, the $\gam$ fusion process.

The $\gam$ production cross--section of heavy leptons, $\gam\rightarrow L^-L^+$, has
been studied in comparison with DY in $pp$ collisions at LHC energies \cite{Drees}.
The full $\gam$ process includes the individual regimes of inelastic,
semi--elastic, and elastic scattering.  The lepton production cross--section in each regime was
found to be of the same order of magnitude while the total $\gam$ cross--section, 
the sum of the individual regimes, was found to be nearly $10^{2}$ below the DY cross--section.  
We repeat the $\gam$ calculations for monopole production which entails 
replacing $e$ (for leptons) with $g\beta$ (for monopoles) that, 
in light of eq.~(\ref{eq:monocharge}), will lead to greatly enhanced cross--sections.
The DY cross--section will also be enhanced for monopole relative to lepton production,
however it remains to be seen which process will dominate.

We restrict our calculations to Dirac monopoles (hereafter ``monopoles''),
assumed to be spin 1/2 fermions of minimal charge, only consider electromagnetic interactions, 
and assume a monopole--photon coupling $g\beta$ for final state monopoles of velocity $\beta.$  
In Section (\ref{sec:twophoton}) we describe the $\gam$ and DY calculations
and discuss their relative dominance for lepton versus monopole production.   
The details of the $\gam$ and DY calculations, which have been widely 
reported in the literature, are presented in Appendix (\ref{sec:appendix})
for the case of monopole production.
In Section (\ref{sec:collider}) we present our results for $p\bar{p}$ collisions
at $\sqrt{s}=1.96~\tev$, discuss the monopole lower mass 
bound reported by CDF, give a mass limit based on our calculations, 
and present our results for $pp$ collisions
at $\sqrt{s}=14~\tev$.  Our concluding remarks are given in Section (\ref{sec:conclusions}).

\section{Photon Fusion Versus The Drell--Yan Process for Monopole Production}
\label{sec:twophoton}

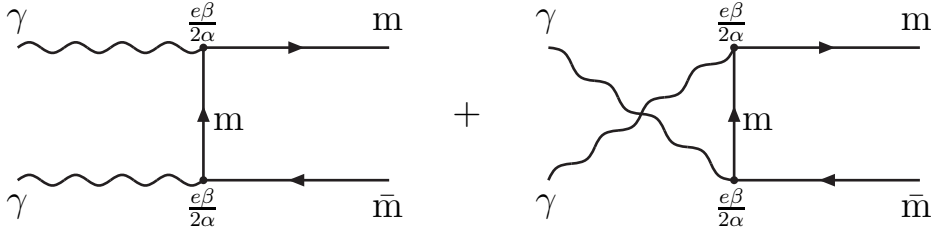
\begin{figure}
\begin{center}
\begin{picture} (400,80)(0,-20)
\SetWidth{1.0}
\SetScale{1.}
\Photon(5,5)(75,5){2}{4}
\Vertex(75,55){1.5}
\Photon(5,55)(75,55){2}{4}
\ArrowLine(145,5)(75,5)
\Vertex(75,5){1.5}
\ArrowLine(75,55)(145,55)
\ArrowLine(75,5)(75,55)
\Text(75,65)[c]{$\frac{e\beta}{2\alpha}$}
\Text(75,-5)[c]{$\frac{e\beta}{2\alpha}$}
\Text(5,65)[c]{{\Large $\gamma$}}
\Text(5,-5)[c]{{\Large $\gamma$}}
\Text(145,65)[c]{{\Large m}}
\Text(145,-5)[c]{{\Large $\bar{\rm{m}}$}}
\Text(85,27)[c]{\Large{m}}
\Text(175,30)[c]{\Large{+}}
\Photon(205,5)(275,55){2}{4}
\Vertex(275,55){1.5}
\Photon(205,55)(275,5){2}{4}
\ArrowLine(345,5)(275,5)
\Vertex(275,5){1.5}
\ArrowLine(275,55)(345,55)
\ArrowLine(275,5)(275,55)
\Text(275,65)[c]{$\frac{e\beta}{2\alpha}$}
\Text(275,-5)[c]{$\frac{e\beta}{2\alpha}$}
\Text(205,65)[c]{{\Large $\gamma$}}
\Text(205,-5)[c]{{\Large $\gamma$}}
\Text(345,65)[c]{{\Large m}}
\Text(345,-5)[c]{{\Large $\bar{\rm{m}}$}}
\Text(285,27)[c]{\Large{m}}
\end{picture}
\end{center}
\caption[Monopole Production via Photon Fusion.]
{The Feynman diagrams for 
the $\gam$ fusion subprocess which produce a monopole-antimonopole
pair ($m\bar{m}$) in the final state.  Two incoming virtual photons ($\gamma$) 
are radiated from the interacting protons or antiprotons (not shown). The virtual photons 
couple to the total charge distribution of the proton (during elastic scattering,
which leaves the proton intact) or to a constituent quark within the proton
(during inelastic scattering).  The monopole--photon coupling is found from 
minimal assumptions of monopole interactions described in the text. 
The $\gam$ cross--section formula for spin 1/2 monopoles is given in eq.~(\ref{crosssec}).} 
\label{fig:2gamma}
\end{figure}

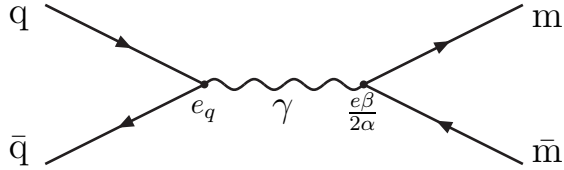
\begin{figure}
\begin{center}
\begin{picture} (200,80)(0,0)
\SetWidth{1.0}
\SetScale{1.}
\ArrowLine(70,40)(10,10)
\ArrowLine(10,70)(70,40)
\Vertex(70,40){1.5}
\Photon(70,40)(130,40){2}{4}
\Vertex(130,40){1.5}
\ArrowLine(130,40)(190,70)
\ArrowLine(190,10)(130,40)
\Text(130,30)[c]{$\frac{e\beta}{2\alpha}$}
\Text(70,30)[c]{$e_q$}
\Text(100,30)[c]{{\Large $\gamma$}}
\Text(0,65)[c]{{\Large q}}
\Text(0,15)[c]{{\Large $\bar{\rm{q}}$}}
\Text(200,65)[c]{{\Large m}}
\Text(200,15)[c]{{\Large $\bar{\rm{m}}$}}
\end{picture}
\end{center}
\caption[Monopole Production via the Drell--Yan Process.]
{The Feynman diagram for 
the Drell--Yan subprocess which produces a monopole-antimonopole
pair ($m\bar{m}$) in the final state.  
An incoming quark and antiquark ($q\bar{q}$), constituants of the
colliding protons, 
annihilate into a virtual photon ($\gamma$) which then pair produces the final
state monopoles. 
The DY cross--section formula for spin 1/2 monopoles is given in eq.~(\ref{eq:DYcrosssec}).} 
\label{fig:DY}
\end{figure}

The $\gam$ and DY processes for lepton production have been widely studied
in proton collisions. 
The $\gam$ subprocess for monopole production, depicted in Fig.~(\ref{fig:2gamma}), 
yields a monopole--antimonopole pair $m\bar{m}$ in the final state. 
The incident photons are radiated from the electric charge distribution
of the colliding protons (or antiprotons).  During elastic scattering the 
photon couples to the whole proton charge $e$ and during inelastic scattering
couples to constituant quarks of charge $e_q=\eta e$ where $\eta=2/3(-1/3)$ 
for $q=u,c,t(d,s,b).$  
We will compare $\gam$ with the DY process (Fig.~(\ref{fig:DY})) which 
dominates for lepton production.  
The full $\gam$ and DY cross--section formulae are presented in Appendix (\ref{sec:appendix}) 
with the relevant couplings for monopole production.

To understand the relative strengths of $\gam$ and DY production it is instructive
to compare the electromagnetic couplings in the total cross--sections. 
For the estimates to follow we only consider quark--photon 
couplings $e_q=\eta e$ where $\eta=2/3(-1/3)$ for $q=u,c,t(d,s,b)$. 
The full $\gam$ calculation includes couplings to the whole proton charge
which will provide a marginal increase to the $\gam$/DY enhancement we find
in eq.~(\ref{eq:ratio}) below. 
In the case of {\it{lepton}} production, $\gam$ suppression relative to DY is anticipated merely
by counting the powers of electromagnetic couplings in the total cross--sections.
The ratio of electromagnetic couplings in the lepton production cross--sections for $\gam$ relative 
to DY is   
\beq
r_l=\frac{e_q^4 e^4}{e_q^2 e^2}=\bar{\eta}^2\alpha^2
\eeq
where $\bar{\eta}$ is the average fractional quark charge contributing to the cross--section.
Going from lepton to monopole production 
we simply replace $e\rightarrow g\beta = e\beta/2\alpha$ in the final state
couplings (as shown in Figs.~(\ref{fig:2gamma}) and (\ref{fig:DY})).  
In this case the ratio of couplings is
\beq
r_m=\frac{e_q^4\left(\frac{e\beta}{2\alpha}\right)^4}
{e_q^2\left(\frac{e\beta}{2\alpha}\right)^2} 
=\bar{\eta}^2\frac{\beta^2}{4}
\eeq
(where $\alpha\equiv e^2$).
We can now estimate the change in the $\gam$/DY cross--section ratio
expected for monopole versus lepton production by taking a ratio of the ratios
\beq
R=\frac{r_m}{r_l}=\frac{\beta^2/4}{\alpha^2}\sim~4700~
\label{eq:ratio}
\eeq
setting $\beta=1.$
Drees, {\it{et al.,}} find $\gam$ production of leptons to be 
nearly $10^2$ below DY \cite{Drees} for $pp$ collisions at 
$\sqrt{s}=14\tev$, which implies a factor $\sim50$ 
{\it{dominance}} of $\gam$ over DY for monopole production 
assuming $\beta=1$ in eq.~(\ref{eq:ratio}).  The effect of
$\beta<1$ for the production of slow moving monopoles is to be determined 
in our full calculation which follows.

\section{Monopole Production in Proton Collisions}
\label{sec:collider}

We calculate $\gam$ fusion for monopoles production 
following the formalism of Drees, {\it{et al.}} \cite{Drees}.  The detailed
formulae are presented in Appendix (\ref{sec:appendix})
and full documentation of our $\gam$ calculations is reported in 
a thesis of one of the authors \cite{Dougall}.  
Unlike the DY process, $\gam$ production yields equivalent 
results in $pp$ and $p\bar{p}$ scattering.  
The full $\gam$ calculation includes contributions from
three individual regimes; inelastic, semi--elastic, and elastic scattering,
and we sum these individual regimes to find the total $\gam$ cross--section.   
For inelastic scattering, 
$pp\rightarrow XX\gamma\gamma \rightarrow XXm\bar{m}$, both intermediate
photons are radiated from partons (quarks or antiquarks) in the colliding protons.  
To approximate the quark distribution within the proton we use the 
Cteq6--1L parton distribution functions \cite{CTEQ} and choose $Q^2=\hat{s}/4$ throughout.
Following \cite{Drees}, we employ an equivalent--photon approximation \cite{WW} for the photon
spectrum of the intermediate quarks. 
In semi--elastic scattering,
$pp\rightarrow pX\gamma\gamma \rightarrow pXm\bar{m}$, one intermediate
photon is radiated from a quark, as in the inelastic process, while the 
second photon is radiated from the other proton,
coupling to the total proton charge and leaving a final state proton intact.
The photon spectrum associated with the interacting proton must be 
altered from the equivalent--photon approximation for quarks to account
for the proton structure.  To accommodate the proton structure we use the modified 
equivalent--photon approximation of \cite{Drees2}. 
For elastic scattering, 
$pp\rightarrow pp\gamma\gamma \rightarrow ppm\bar{m}$, both intermediate
photons are radiated from the interacting protons leaving both protons intact 
in the final state.

\begin{figure}[t]
\includegraphics[scale=0.5]{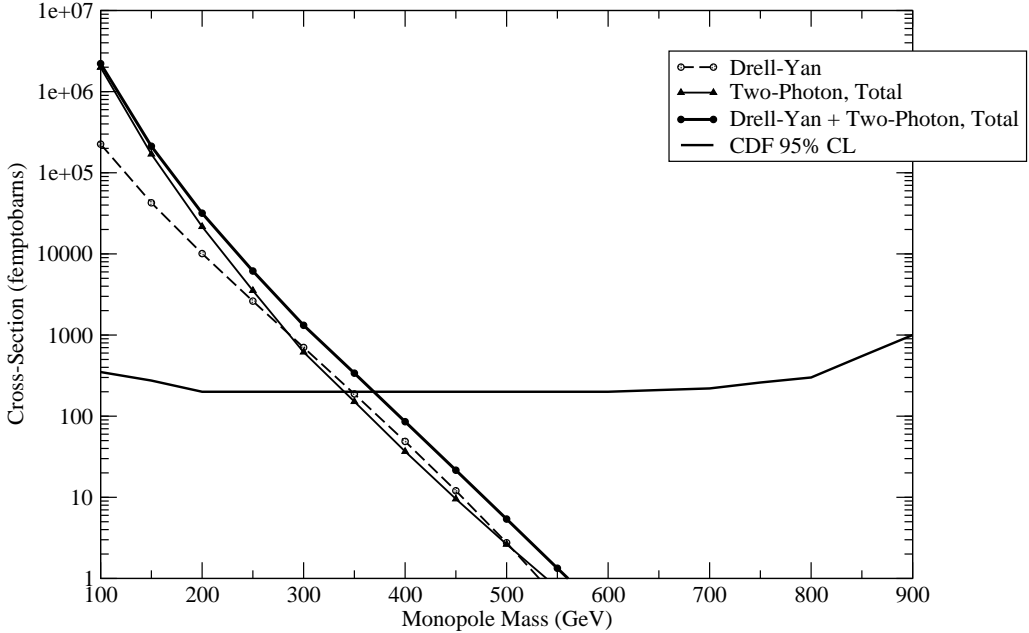}
\caption{\label{Tevatron}
The total cross--sections for $\gam$ and DY 
production of monopoles in $p\bar{p}$ scattering vs.~the mass of the produced 
monopole at $\sqrt{s}=1.96$~TeV.  
The CDF reported cross--section limit at 95\% confidence level (shown above) 
can be used to set a lower monopole mass limit assuming a production mechanism.  
The $\gam$ and DY curves are nearly equal 
at the exclusion limit and independently call for a mass limits $m>345\gev$
and $m>350\gev$, respectively.
We also plot the sum of $\gam$ and DY and find a monopole mass limit
$m>370\gev$ assuming both production mechanisms.
}
\end{figure}

The Fermilab Tevatron is a $p\bar{p}$ collider at center--of--mass energy
$\sqrt{s}=1.96\tev.$  We have calculated $\gam$ and DY production of monopoles
at the Tevatron and our results are presented in Fig.~(\ref{Tevatron}). 
The individual $\gam$ scattering regimes are all of similar magnitude and
happen to fall near the DY curve.  To avoid cluttering Fig.~(\ref{Tevatron})
the individual regimes are not shown, but their
contributions to the total $\gam$ cross--section are roughly:
inelastic (10\%),  semi--elastic (50\%), and elastic (40\%).
The total $\gam$ and DY cross--sections shown in Fig.~(\ref{Tevatron})  
happen to be nearly equal over the range $300$ to $500~\gev$ with $\gam$ dominating 
at lower masses.

The result of a search for the direct production of monopoles 
was recently reported by the CDF Collaboration \cite{CDF06}.  
The search  uses $35.7~\rm{pb}^{-1}$ of CDF run II data where a special monopole 
trigger was employed.  Monopole event simulations and conservative 
estimates of their experimental acceptance were used to establish a cross--section 
limit of approximately 200 femptobarns over a mass range $200$ to $700~\gev$ 
at 95\% confidence level based upon a lack of observed monopole events (see 
Fig.~(\ref{Tevatron})). 
Assuming DY production of monopoles, CDF establishes a monopole mass  limit of $m>360\gev$.  
Their monopole acceptance depends upon the production kinematics, but they estimate
a limit in the total variation in acceptance to be less than 10\% and conclude that mass
limits from production mechanisms other than DY can be set with reasonable accuracy.   
Thus, we are justified in considering mass limits from $\gam$ production based
upon the CDF 95\%CL limit.

The DY curve shown in Fig.~(3) of \cite{CDF06} crosses the 95\% CL limit near $360\gev$
while our DY curve is slightly lower and crosses near $350~\gev.$
Therefore, our DY calculation calls for a slightly lower mass limit than CDF reports, 
but the addition of the $\gam$ contribution to the DY production 
argues for an increase in the mass limit of $20\gev$ to $m>370\gev.$ 
See the $\gam$+DY curve in Fig.~(\ref{Tevatron}).

\begin{figure}[t]
\includegraphics[scale=0.5]{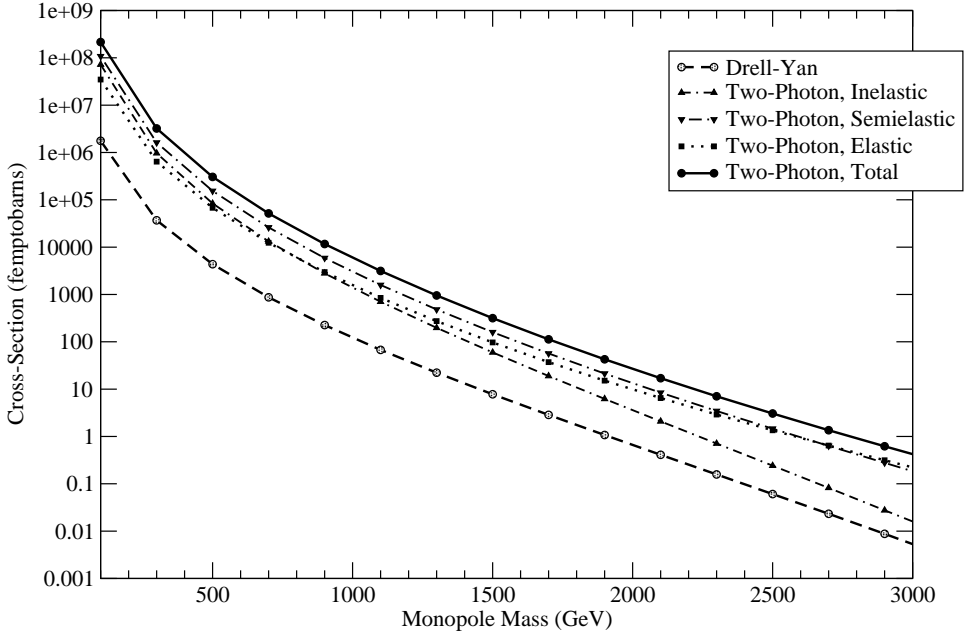}
\caption{\label{LHC}
The cross--sections for $\gam$ fusion and DY 
processes in a $pp$ interaction vs. the mass of the produced 
monopole at $\sqrt{s}=14$~TeV.  
The $\gam$ process is calculated in the elastic, semi--elastic and inelastic 
regimes shown above and the sum of the individual processes gives the total $\gam$ 
production cross--section (solid line). }
\end{figure}

The LHC is designed to produce $pp$ collisions copiously at $\sqrt{s}=14$~TeV.
The results of our calculations at LHC are presented in Fig.~(\ref{LHC}).  
We find that each of the individual $\gam$ scattering regimes dominates DY by a factor $> 10$ and  
the total $\gam$ cross--section is a factor $> 50$ larger than DY.   
Based upon our results we conclude that $\gam$ fusion 
will be the leading mechanism for direct monopole production at LHC
and argue for further investigation of the $\gam$ process 
in detailed simulations of LHC monopole events.  

If the LHC were to attain $100~\rm{fb}^{-1}$ of integrated luminosity our calculations
predict greater than 700,000 monopole events from $\gam$ fusion for a monopole
mass of $1\tev.$  By comparison, the yield of $1\tev$ monopoles from DY production is less
than 15,000 events over the same period of time.  The $\gam$ process will allow LHC to
extend their monopole search to relatively high masses.  After collecting
$100~\rm{fb}^{-1}$ of data we predict in excess of 50 monopole events at
monopole masses approaching $3\tev.$

\section{Conclusions}
\label{sec:conclusions}

Motivated by recent monopole searches at the Fermilab Tevatron
and the expectation of future monopole searches at the LHC
we have calculated monopole production from the Drell--Yan and $\gam$ fusion processes.   
We have compared these processes for both $p\bar{p}$ collsions at
$\sqrt{s}=1.96\tev$ and $pp$ collisons at $\sqrt{s}=14\tev$.
Our calculations are limited to a lowest order estimate assuming
a monopole--photon coupling proportional to the induced electric
field of a moving monopole.  The monopole--photon coupling is
strong for monopole velocities $\beta\sim 1$ which has prohibited our use of a 
pertubative expansion.    

In $p\bar{p}$ collisions
at the Tevatron we showed that the $\gam$ total cross--section
is approximately equal to DY in the mass range where the CDF collaboration
sets a 95\% confidence level cross--section limit. 
Based on our results, DY can be used set a mass limit $m>350\gev$
and $\gam$ can be used independently to set a mass limit $m>345\gev$.  
When both $\gam$ and DY production are considered, 
the sum of the cross--sections implies the monopole mass limit of $m>370\gev$.
   
In $pp$ collisions
at the LHC we found that $\gam$ fusion is the dominant production mechanism
for magnetic monopoles by more than a factor 50 over the DY process.
The inelastic, semi--elastic, and elastic regimes each dominate DY by 
a factor 10 or greater.    
We conclude that the $\gam$ process should be considered 
the leading production mechanism for monopole searches at the LHC
and emphasize the need for detailed studies of monpole events using 
the $\gam$ process.

\appendix
\section{Appendix}
\label{sec:appendix}

Our $\gam$ calculations follow the formalism and approximations of
Drees {\it{et al.,}} \cite{Drees}.  We calculate inelastic, semi--elastic,
and elastic processes and assume throughout this report 
that the final state monopoles are Dirac--type of minimal charged
($n=1$), spin 1/2 fermions, and only consider their electromagnetic couplings.    
The $\gam$ subprocess must satisfy
the kinematic constraint $\hat{s}=(k_1+k_2)^2 \geq 4m^2$ where $k_1$ 
and $k_2$ are the virtual photon four--momenta
and the final state monopole pair has a total rest mass $2m$.   
We assume an effective photon approximation \cite{WW} to describe the photon spectrum
of the interacting quark during inelastic scattering.  The total cross--section
for inelastic scattering is
\bea 
\sigma_{pp}^{inel.}(s)&=& \sum_{q,\; q'} \int_{4m^2/s}^1dx_1
\int_{4m^2/sx_1}^1 dx_2 \int_{4m^2/sx_1 x_2}^1 dz_1
\int_{4m^2/sx_1 x_2 z_1}^1 dz_2 \;\;e_q^2 e_{q'}^2 \nonumber \\
& \cdot &f_{q/p}(x_1,\;Q^2)\;
f_{q'/p}(x_2,\; Q^2)f_{\gamma
/q}(z_1)\;f_{\gamma /q'}(z_2) \;\hat{\sigma}_{\gamma \gamma}(x_1 x_2 z_1 z_2s)
\label{inelast}
\eea
where $m$ is the monopole mass, 
$e_q=\eta e$ where $\eta=2/3(-1/3)$ for $q=u,c,t(d,s,b)$, 
and $\hat{\sigma}_{\gamma \gamma}$ is the production
subprocess cross--section with the center--of--mass energy
$\sqrt{\hat{s}} = \sqrt{x_1 x_2 z_1 z_2s}$.  The structure function
$f_{q/p}$ is the quark density  inside the proton and $f_{\gamma / q}$ is 
the equivalent--photon spectrum of a quark. We use the Cteq6-1L parameterization of the parton
densities \cite{CTEQ} and chose the scale $Q^2=\hat{s}/4$. 
With
\beq 
f_{\gamma /q}(z)=f_{\gamma /q'}(z)
= \frac{\alpha}{2 \pi}\;\;\frac{(1+(1-z)^2)}{z}\;\;\ln(Q_{\rm{max}}^2/Q_{\rm{min}}^2)
\label{eq:WW}
\eeq
where $Q^2_{\rm{max}}=\hat{s}/4-m^2 $ and  $Q^2_{\rm{min}}=1\gev^2$.

The final state monopole velocity is $\beta=(1 -4m^2/\hat s)^{1/2}$ for the subprocess 
center--of--mass energy $\sqrt{\hat{s}}$.  The $\gamma\gamma\rightarrow m\bar{m}$
total cross--section is
\beq 
\hat{\sigma}(\gamma \gamma \to m \bar{m})= 
\frac{\pi\beta^5}{4\alpha^2\hat{s}}
\left[\frac{3-\beta^4}{2\beta} 
\ln \frac{1 + \beta}{1 -\beta}
-(2-\beta^2)\right]. 
\label{crosssec}
\eeq
where $g^4\beta^4=\beta^4/16\alpha^2$ using the Dirac quantization condition.  
The factor $\alpha^{-2}$ will be cancelled by two powers of $\alpha$ from
eqs.~(\ref{eq:WW}) and (\ref{eq:WWp}).

The semi-elastic cross section for $pp \to m\bar{m}pX$ is given by
\bea 
\sigma_{pp}^{semi-el.}(s)&=& 2\; \sum_q 
\int_{4m^2/s}^1dx_1\int_{4m^2/sx_1}^1dz_1
\int_{4m^2 /sx_1 z_1}^1dz_2\;e_{q}^{2}\;f_{q/p}(x_1,\;Q^2)\; \nonumber \\
& \cdot & f_{\gamma}(z_1)
\;f_{\gamma /p}^{el.}(z_2)\;\hat{\sigma}_{\gamma \gamma}(x_1 z_1
z_2 s) 
\label{semielast}
\eea
The subprocess energy now is given by $\sqrt{\hat s} = \sqrt{sx_1 z_1z_2} $.

For the elastic photon spectrum $f_{\gamma /p}^{el.}(z)$ we use an analytic
expression from \cite{Drees2} given by 
\beq 
f_{\gamma /p}^{el.}(z)= \frac{\alpha}{2 \pi z}
(1+(1-z)^2)\left[\ln A - \frac{11}{6}+\frac{3}{A}-\frac{3}{2A^2}
+\frac{1}{3A^3}\right]~~,
\label{eq:WWp}
\eeq
for
\beq 
A=1+\frac{0.71({\rm GeV})^2}{Q_{\rm{min}}^2}~~, 
\eeq
and where
\bea 
Q_{\rm{min}}^2&=&-2m_p^2 + \frac{1}{2s}\biggl[(s+m_p^2)(s-zs+m_p^2)
\nonumber \\
&-& (s-m_p^2)
\sqrt{(s-zs-m_p^2)^2- 4m_p^2zs}\biggr]~~.
\eea
At high energies $Q_{\rm{min}}^2$ is approximately
$m_p^2 z^2/(1-z)$.

The purely elastic scattering cross--section where both protons
remain intact in the final state is
\beq
\sigma_{pp}^{el.}(s)=\int_{4m^2/s}^1 dz_1 \int_{4m^2/z_1s}^1
dz_2 \;\;f_{\gamma /p}^{el.}(z_1)\;f_{\gamma /p}^{el.}(z_2)\;
\hat{\sigma}_{\gamma \gamma}(\hat s = z_1z_2s). 
\label{elast}
\eeq

In the DY process the annihilating $q\bar{q}$ pair
must satisfy $\hat{s}=(p_1+p_2)^2 \geq 4m^2$, for quark four--momenta 
$p_1$ and $p_2$, to produce a final state monopole pair of total rest mass $2m$.   
The Drell--Yan cross--section for monopole production is
\beq
\sigma_{pp}^{\rm{DY}}(s) \,= \, \sum_{q}\int_{4m^2/s}^1 dx_1 \int_{4m^2/x_1s}^1
dx_2 \;\;f_{q /p}(x_1)\;f_{\bar{q} /p}(x_2)\;
\hat{\sigma}_{q\bar{q}}(\hat s = x_1x_2s) 
\label{eq:DY}
\eeq
for the DY subprocess  
\beq 
\hat{\sigma}(q \bar{q} \to m \bar{m})= 
\frac{\pi \eta^2 \beta^3}{12\hat{s}}
\left[2-\frac{2}{3}\beta^2\right] 
\label{eq:DYcrosssec}
\eeq
where $\eta$ is the fractional quark charge in units of $e$ and 
the quark sum ranges from $\bar{t},\bar{b},...,b,t$, ensuring that
only quarks and antiquarks of the same flavor contribute.


\acknowledgments

T.~D.~acknowledges research support from SMU's Dedman College.
Both authors acknowledge useful discussions with 
Randall Scalise, Ryzsard Stroynowski, and Jingbo Ye.

\end{document}